\documentclass[review,authoryear]{elsarticle}

\usepackage[labelfont=bf,]{caption}
\usepackage{hyperref}

\usepackage{ulem}

\usepackage{color}

\journal{Astronomy and Computing}

\biboptions{authoryear}

\begin{document}

\begin{frontmatter}

\title{Towards an Astronomical Science Platform: \newline
    Experiences and Lessons Learned from Chinese Virtual Observatory}
\tnotetext[mytitlenote]{The Chinese Virtual Observatory cloud platform can be accessed at: \href{http://astrocloud.china-vo.org/}{http://astrocloud.china-vo.org/}}

\author[NAOC]{Chenzhou Cui \corref{correspondingauthor}}
\cortext[correspondingauthor]{Corresponding author}
\ead{ccz@bao.ac.cn}
\author[NAOC]{Yihan Tao}
\author[NAOC]{Changhua Li}
\author[NAOC]{Dongwei Fan}
\author[TJU]{ Jian Xiao}
\author[NAOC]{Boliang He}
\author[NAOC]{Shanshan Li}
\author[TJU]{Ce Yu}
\author[NAOC]{Linying Mi}
\author[NAOC]{Yunfei Xu}
\author[NAOC]{Jun Han}
\author[NAOC]{Sisi Yang}
\author[NAOC]{Yongheng Zhao}
\author[NAOC]{ Yanjie Xue}
\author[NAOC]{Jinxin Hao}
\author[PMO]{Liang Liu}
\author[SHAO]{Xiao Chen}
\author[YNAO]{Junyi Chen}
\author[XAO]{Hailong Zhang}

\address[NAOC]{National Astronomical Observatories, Chinese Academy of Sciences, 20A Datun Road, Chaoyang District, Beijing 100101, China}
\address[TJU]{College of Intelligence and Computing, Tianjin University, No. 135 Yaguan Road, Haihe Education Park, Tianjin 300350, China}
\address[PMO]{Purple Mountain Observatory, Chinese Academy of Sciences, No. 10 Yuanhua Road, Qixia District, Nanjing 210033,China}
\address[SHAO]{Shanghai Astronomical Observatory, Chinese Academy of Sciences, 80 Nandan Road, Shanghai 200030, China}
\address[YNAO]{Yunnan Observatories, Chinese Academy of Sciences, No. 396 Yangfangwang, Guandu District, P.0. Box 110, Kunming 650011, Yunnan, China}
\address[XAO]{Xinjiang Astronomical Observatory, Chinese Academy of Sciences, 150 Science 1-Street, Urumqi, Xinjiang 830011, China}

%


\begin{abstract}
In the era of big data astronomy, next generation telescopes and large sky surveys produce data sets at the TB or even PB level. Due to their large data volumes, these astronomical data sets are extremely difficult to transfer and analyze using personal computers or small clusters. In order to offer better access to data, data centers now generally provide online science platforms that enable analysis close to the data. The Chinese Virtual Observatory (China-VO) is one of the member projects in the International Virtual Observatory Alliance and it is dedicated to providing a research and education environment where globally distributed astronomy archives are simple to find, access, and interoperate. In this study, we summarize highlights of the work conducted at the China-VO, as well the experiences and lessons learned during the full life-cycle management of astronomical data. Finally, We discuss the challenges and future trends for astronomical science platforms.
\end{abstract}

\begin{keyword}
big data \sep cloud computing \sep science platform \sep virtual observatory
\end{keyword}
\end{frontmatter}

\section{Introduction}
Huge amounts of astronomical data have been generated due to the development of large ground-based and space-based telescopes as well as digital sky surveys. Current sky surveys such as the Sloan Digital Sky Survey (SDSS) \citep{2000AJ....120.1579Y}, Large Sky Area Multi-Object Fiber Spectroscopic Telescope (LAMOST) \citep{2012RAA....12.1197C}, Gaia \citep{2016A&A...595A...1G}, and Transiting Exoplanet Survey Satellite \citep{2015Ricker_TESS} have produced hundreds of TB of data. In the 2020s, more data will be produced at the PB and even EB level by several next-generation large telescopes and surveys around the world, e.g., SKA \citep{2009IEEEP..97.1482D}, FAST \citep{2011IJMPD..20..989N}, LSST \citep{2019ApJ...873..111I}, Euclid\citep{2011arXiv1110.3193L}, and WFIRST\citep{2012arXiv1208.4012G}. Thus, astronomy is entering a new era of big data where the data sets are too large for astronomers to download and analyze using their own computers, thereby calling for the development of an online science platform for astronomy \citep{Zhang2016067,2019BAAS...51g.146D}.

Astronomy has always had a tradition of making data open and several pioneers have provided online data services and tools. For example, the Strasbourg Astronomical Data Centre (CDS)\citep{2000A&AS..143....1G} established in 1972 has been dedicated to the dissemination of value-added digital astronomical data. The CDS developed a series of well-known tools for astronomers to allow them to query, explore, and analyze data online, including SIMBAD\citep{2000A&AS..143....9W}, VizieR\citep{2000A&AS..143...23O}, and Aladin\citep{2000A&AS..143...33B}. The SDSS data release system is another example. The browser- and SQL-based interfaces provided by SkyServer and CasJobs \citep{2005cs........2072O} were designed to allow astronomers to access and filter SDSS data in a convenient manner. These online data services and tools which make data easily discoverable laid a solid groundwork for the science platform idea.

A virtual observatory (VO) is a data-intensive online astronomical research and education environment based on advanced information technologies for achieving seamless and global access to astronomical information \citep{2008IAUS..248..563C}. In order to work toward the same goal, the International Virtual Observatory Alliance (IVOA) was established in 2002 with the aim of connecting worldwide data centers and projects by building a framework for discussing and sharing VO ideas and technology\citep{2012epsc.conf..626A}. In China, the Chinese Virtual Observatory (China-VO) was established in the same year as IVOA\citep{2004PNAOC...1..203C}. Furthermore, the Chinese Astronomical Data Center was established as a member of the World Data Center system in 1989. During the last decade,  China-VO has developed Astrocloud, a science platform that integrates data, tools, and computing resources for astronomical observations, research, and education. The Astrocloud platform is open to all users, including professional astronomers, amateur astronomers, students, educators and the public all over the globe.

\section{Full Life-cycle Management of Astronomical Data}
\begin{figure}[h]
    \centering
    \includegraphics[width=0.8\textwidth]{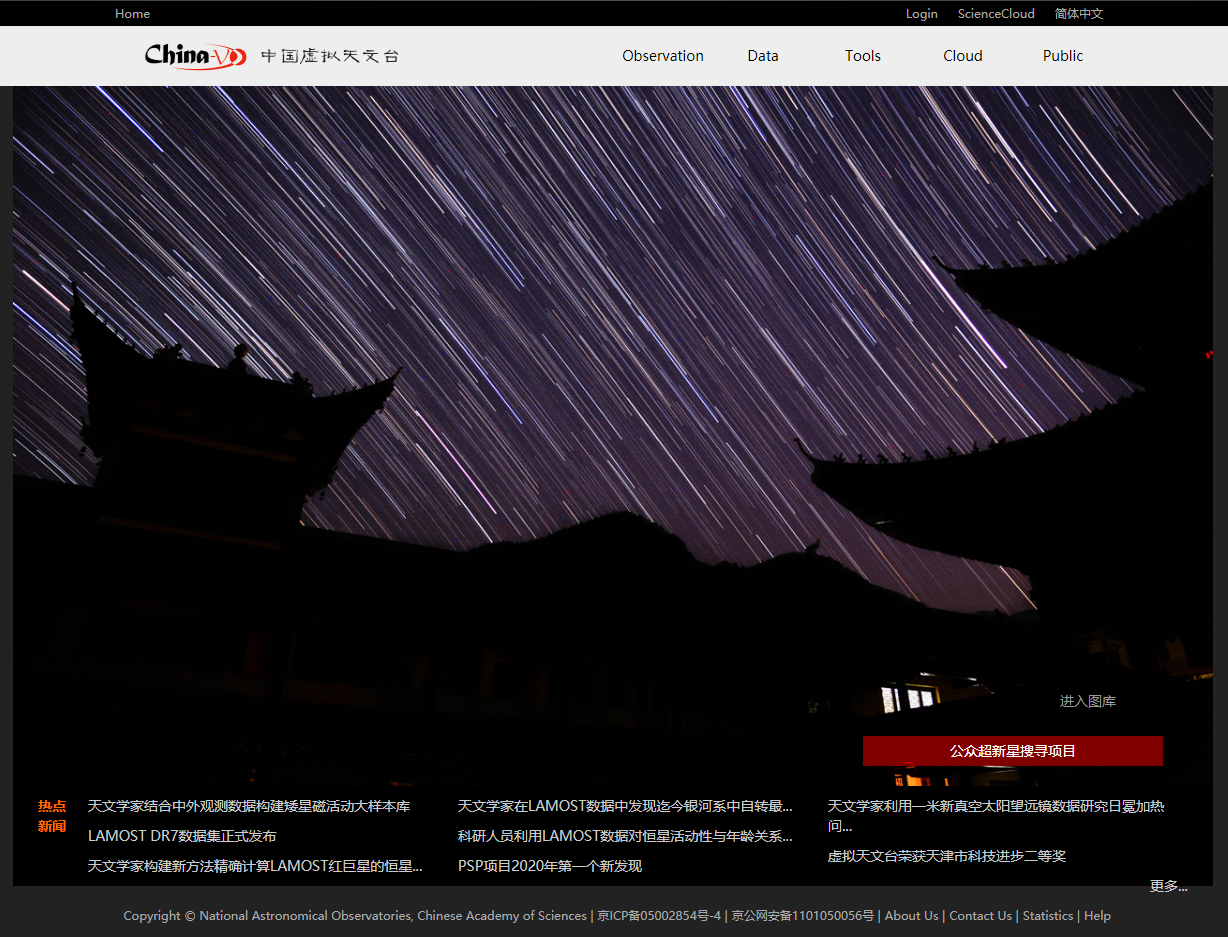}
    \caption{China-VO AstroCloud portal.}
    \label{fig:portal}
\end{figure}

China-VO developed AstroCloud as cyber-infrastructure for astronomy research and public outreach, with the aim of facilitating full life-cycle management of astronomical data, including observation proposal submission, data archiving, data release and open access, cloud computing for data processing and analysis, and data-related projects for the public \citep{2017arXiv170105641C}. Fig. \ref{fig:portal} shows the portal of the AstroCloud platform.

 By the end of 2019,  Astrocloud served more than 20,000 registered users, including more than 3100 professional astronomers, and it received over 10 million visits, serviced nearly 4.1 million searches and more than 20 million times of scientific data downloads. The platform provides full life-cycle data services for LAMOST, FAST and other national key astronomical observation equipments. It also provided a cloud computing environment for graduate teaching at the University of the Chinese Academy of Sciences and National Astronomical Observatories, Chinese Academy of Science (NAOC).

\begin{figure}[h]
    \centering
    \includegraphics[width = \textwidth]{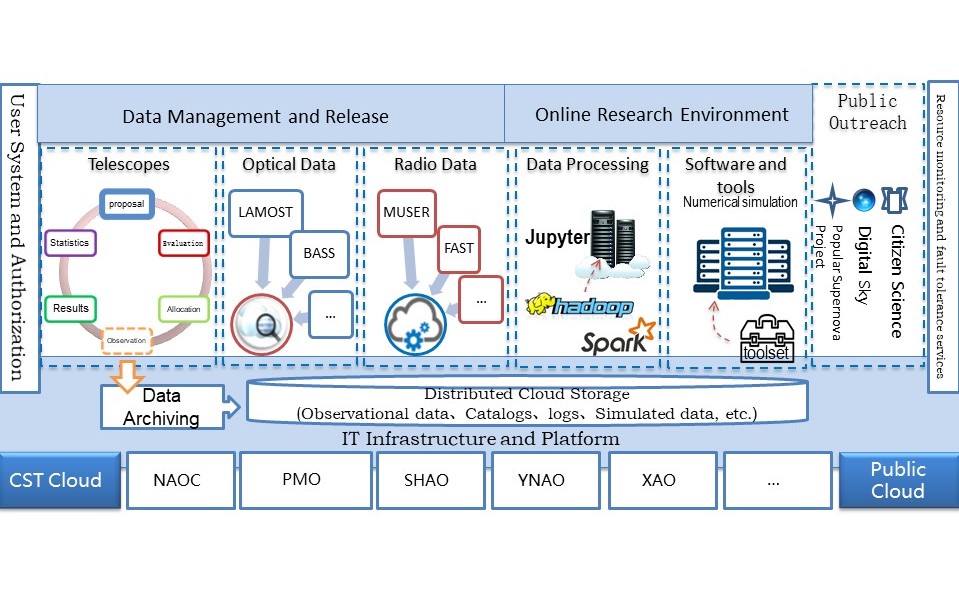}
    \caption{Architecture of China-VO AstroCloud platform.}
    \label{fig:astrocloud_architecture}
\end{figure}

The AstroCloud platform was developed based on the CloudStack\footnote{http://cloudstack.apache.org/} middleware environment by integrating third-party cloud resources, i.e., CSTCloud \footnote{http://www.cstcloud.cn/} and Alibaba Cloud \footnote{https://www.alibabacloud.com/}. As shown in Fig.\ref{fig:astrocloud_architecture}, five leading astronomical observatories in China serve as distributed nodes and the platform is built on a distributed cloud storage layer. Distributed storage, data, and computing resources are managed uniformly by a cloud environment management platform employing virtualization and cloud computing technology in order to provide the upper-level integrated service platforms. User system and authorization, resource monitoring, and fault-tolerant services comprise the control centres of the system, which collect user operation logs and runtime information for each service platform, coordinate the use of system resources, and ensure the stable operation of the system.
\subsection{Telescope Proposal Submission and Peer Review}
For observational astronomers, the first step that is usually required to obtain observational data for their research is applying for telescope time. Since 2014, the AstroCloud platform has supported online proposal submissions and peer review for a number of telescopes in China, including the Lijiang 2.4 m telescope, Xinglong 2.16 m telescope, and the New Vacuum Solar Telescope at Fuxian Lake Solar Observatory. In 2018, China-VO upgraded the proposal submission system to support the creation, editing, and submission of proposals to several overseas telescopes via the CAMS-CAS \footnote{www.cams-cas.ac.cn} leading Telescope Access Program (TAP). TAP provides all China-based astronomers with access to internationally competitive medium- and large-aperture optical-infrared telescope facilities, including the Canada--France--Hawaii Telescope (CFHT), Palomar Hale Telescope, and Las Cumbres Observatory (LCOGT). By September 2019, the AstroCloud platform had received 666 applications from 285 users at more than 60 institutes.

The AstroCloud platform is beneficial for both telescope managers and observers because it allows the whole proposal submission and peer review process to be performed online. the telescope proposal management process comprises four simple sequential phases, namely proposal submission, proposal review, time allocation, and notification and confirmation. The workflow engine automatically drives the whole process. Each phase can be customized to meet the specific requirements of a telescope. In order for a telescope to join the platform, developers only need to create the necessary templates and write a few lines of code for a telescope-specific data model and its validation check. A telescope usually has its own dedicated control system and limited local storage, so the AstroCloud platform also provides a common data transfer interface as a bridge that connects individual telescopes and the telescope proposal platform, which allows a daemon process to periodically move locally observed data into the China-VO cloud storage \citep{Xiao2018}.

\subsection{Data Archiving and Management}
Data interoperability and conforming with the Findable, Accessible,Interoperable, Reusable (FAIR) principle \citep{2016NatSD...360018W} are the key goals of the VO. Data archiving and management is an essential phase in order to achieve the goals of the VO. A VO platform should provide technical support and services for data processing teams, including data transfer, storage and computing, collecting and managing raw observational data and data products.

China-VO implemented an astronomical data archiving system \citep{2015ASPC..495..483H} for archiving astronomical observation data, which features data standardization and quality control. Metadata are stored together with observational data in the database.

\begin{figure}[h]
    \centering
    \includegraphics[width = \textwidth]{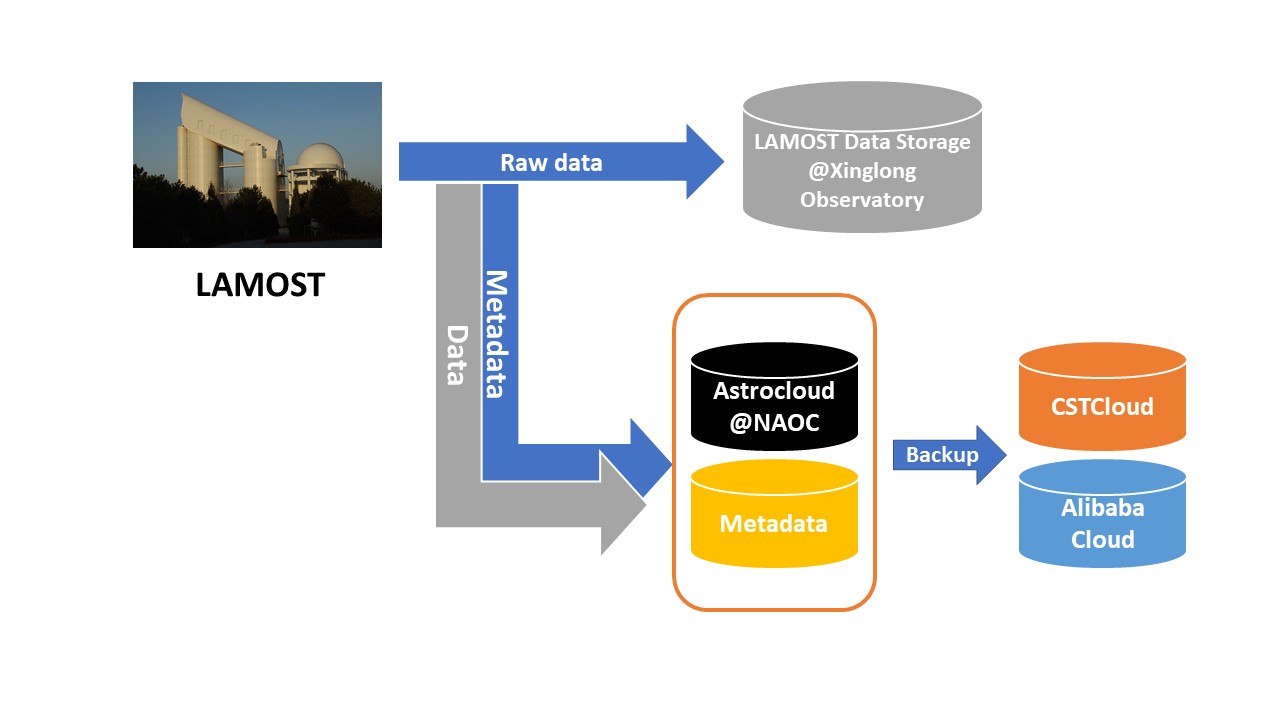}
    \caption{Data archiving process for LAMOST observational data.}
    \label{fig:LAMOST_archiving}
\end{figure}

Astrocloud provides archiving services for telescopes located at multiple sites throughout China. For example, the LAMOST or Guo Shoujing telescope is taking the lead in large-scale and wide-field optical spectral sky surveys, and it has the highest spectral acquisition rate in the world. LAMOST can accomodate 4000 fibers on its focal surface,  by which the collected light of distant and faint celestial objects down to 20.5 magnitudes is fed into the spectrographs, promising a very high spectrum acquiring rate of several ten-thousands of spectra per night.  As shown in Fig.\ref{fig:LAMOST_archiving}, while the LAMOST telescope is operated at Xinglong Observatory according to its observation plan, the observational data and metadata are transmitted automatically each day to the NAOC data center, and backed up in CSTCloud or Alibaba Cloud. The LAMOST data processing team can then access the raw data at the data center in order to process and transfer the data produced back to the data center according to data release plan. 

\subsection{Data Access and Interoperability}
The data access and interoperability sub-system provides a bridge that connects data resources on the cloud with the user. Archived data can be searched using this system and transferred to the cloud storage, which is connected to virtual machines. Thus, astronomers can handle their data on the fly and they do not have to download huge amounts of data to a local machine.

The data access system provides an authorization mechanism to protect data that have not yet been published. Typically, observational data should be released afters 18 months. Before public release, only authorized users can access the data, e.g., the data owner or cooperators. In some cases, a research study may take many years and the data need to be protected until the results are published. The system allows the data owner to set the proprietary period of their data.

The data access system supports many IVOA protocols to facilitate interoperability with other software. For example, using the IVOA-SAMP protocol, a single button click allows the search results to be sent  via sampjs to Topcat, Aladin, or other software that support SAMP protocol. Thus, the web browser and related software can work together in a similar manner to unique local desktop software. The ConeSearch, SSAP and HiPS protocols are supported, and registered in the IVOA registries. Users can employ the service or their own program to access the data and they might not even know which service they are using.

Many other functions are provided to users  as web applications which can be accessed through the AstroCloud website, including online visualization, tables, joined search, and batch download, which support data publishing for LAMOST, SCUSS, AST3, CSTAR, BASS, GMG2.4m, and other projects. Some popular data sets generated by international projects are also mirrored and provided by the system, including TWOMASS, SDSS, Gaia, and UCAC. Increasing amounts of data are being collected and published on this platform to support astronomical research, as well as education and public outreach (EPO).

\section{Highlights of China-VO}
In recent decades, China-VO and Chinese Astronomical Data Center have developed several featured services and user cases for the full life-cycle management of astronomical data. PaperData service is designed for researchers to use VOSpace \citep{2018ivoa.spec.0621G} by employing the data analysis process to publish a scientific paper, where authors benefit from a digital object identifier (DOI) and other identifier references. The AstroCloud platform was introduced into university courses by providing virtual machines and templates that allow lecturers to share their software environments with students. Astronomical data are also processed by China-VO as data sets for data science contests, data-driven astronomy education and public outreach (DAEPO) activities such as citizen science projects, WorldWide Telescope(WWT) teacher training, and WWT Guided Tour \footnote{WorldWide Telescope guided tours are interactive paths through the night sky, designed to tell a story or teach a particular astronomy concept.} competitions .

\subsection{PaperData and DOI}
To conform with the FAIR principles, data resources should have resource identifiers, which link to standard and unique descriptions of the data resources. China-VO uses the standard resource identifier proposed by the IVOA, e.g., the identifier for LAMOST data release one is ivo://China-VO/data/lamost/dr1. In addition, an international uniform DOI will be assigned for data resources such as data sets and paper data.

China-VO provides PaperData with integrated DOI services. PaperData is a VOSpace that astronomers can use to upload and long-term store their data related to a scientific paper, including data sets, tables, figures, pictures, movies, software, models, and source codes. In PaperData, a DOI is employed as a persistent identifier, together with a user-specified URL and an IVO ID, which links to the persistent data repository maintained by the China-VO.

\begin{figure}[h]
    \centering
    \includegraphics[width =0.9\textwidth]{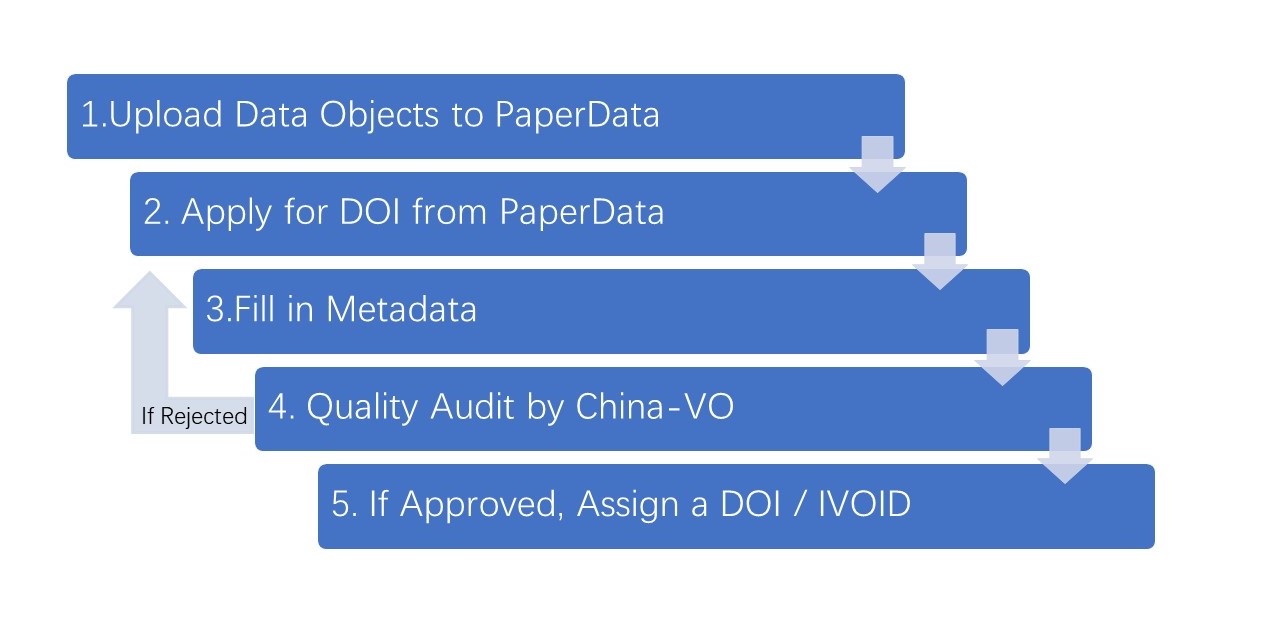}
    \caption{PaperData DOI application procedure.}
    \label{fig:doi}
\end{figure}

The procedure followed to apply for a DOI is shown in Fig.\ref{fig:doi}. Users can apply for a DOI for their uploaded data objects via the PaperData Interface. After completing the related metadata for the data objects, users can submit an application and wait for approval. A DOI will be assigned to the data objects when the paper has been accepted and all the required metadata is provided. In addition, PaperData supports DOI versioning, where a different DOI number will be assigned to a new version of the data object.
\subsection{Cloud Computing, Teaching, and Training}
The China-VO AstroCloud is a cloud platform that allows users to create a virtual machine for data analysis. It also provides virtual machine templates for university courses and summer schools. For example, China-VO provides a standard template for the “Multi-wavelength astronomical data acquisition and processing" course by integrating various types of software, including Heasoft, IRAF, DS9, JHelioview, Python, and IDL, which are used to analyze multi-wavelength data during the course. Students can create a virtual machine with the necessary software in a few minutes by using the templates on the platform. Thus, students can save the time required to install software and configure environments, thereby allowing them to concentrate on the data analysis process. Similar templates are also provided for other courses.

On the other hand, China-VO collaborates with data science competition platforms, such as Alibaba Cloud TianChi, Future Lab, and Kaggle Days, by using real observational astronomical data sets for data mining algorithm competitions. From February to May 2018, a spectral classification algorithm competition was organized by China-VO and Alibaba Cloud. Over 800 teams from all over the world participated in the competition, where they applied machine learning techniques to the LAMOST spectra data set to automatically classify the spectra into star, galaxy, quasar, and unknown objects. The winning team employed deep neural networks to obtain solutions and they achieved an overall macro-F1 score of 0.83. These activities are new attempts by China-VO to form bridges between astronomers and the data science community, and for combining astronomy research with public outreach, thereby obtaining benefits for both astronomers and the public.

\subsection{DAEPO}
China-VO plays a leading role in the IAU DAEPO Working Group, which was officially launched in April 2017. This working group acts as a forum to discuss the value of astronomical data in EPO, the advantages and benefits of data-driven EPO, and the challenges facing DAEPO. It also provides guidelines, curricula, data resources, tools, e-infrastructure, and best practices for DAEPO \citep{2018arXiv180105098C}.

In fact, China-VO has been one of the pioneers of DAEPO by utilizing astronomical data and VO technology to conduct astronomical science education activities and projects for decades.The AstroCloud platform was mainly designed for professional astronomy researchers but it also provides channels that allow the public and amateur astronomers to interact with astronomical data.

For example, in 2015, China-VO started the Popular Supernova Project in collaboration with Xingming Amateur Astronomical Observatory. The image data observed by Xingming Observatory are published online through AstroCloud platform for finding new supernovas. The project aimed to give anyone interested in searching for new supernova objects the opportunity to participate in professional astronomical discoveries. By October 2019, 27 new supernova and extragalactic nova candidates were found, among which 15 supernova and four extragalactic novae were confirmed by professional observations. This was the first citizen science project developed in China based on amateur observational data, and it was also a successful attempt at in depth cooperation between professional and amateur astronomer teams.

Another successful project conducted by China-VO involved the localization of the American Astronomical Society (AAS) WorldWide Telescope (WWT) for Chinese - speaking users, and a series of DAEPO activities using the localized version. The WWT is a powerful visualization tool for demonstrating and exploring astronomical data \citep{2018ApJS..236...22R}. China-VO localized the AAS WWT and established an ecosystem for the WWT, as well as implementing the IVOA HiPS data standard and adding data resources obtained in China to the WWT system. China-VO organized a series of DEAPO activities to promote the usage of WWT in astronomy education, including teacher training and WWT Guided Tours contests. Large volumes of real data are presented to users through the WWT platform. By utilizing the advantages of WWT, China-VO explored user cases and scenarios for DEAPO, such as planetarium, classroom, dome, and VR interaction examples.

\section{Lessons Learned and Challenges}
China-VO has learned several lessons based on the experience of developing AstroCloud and supporting the full life cycle of astronomical data management. Two key challenges still need to be addressed in order to operate a global science platform network in the next decade.

\subsection{Single Sign-On (SSO) and Interconnection among Different Science Platforms}
IVOA advocates SSO authentication among VO projects, which means that a user can sign on at a single point and access all the trustworthy archives, storage, and processing facilities integrated in the entire VO. The interconnection and mutual authorization among different VO projects and science platforms is very important. The platforms at different data centers and VO projects should be connected to allow cross-platform analysis and processing \citep{2019BAAS...51g.146D} by relying on a SSO system.

China-VO understands the importance of SSO authentication and the current platform still needs to be improved to achieve SSO. The China-VO AstroCloud platform uses the CSTCloud passport for authentication. After signing on to AstroCloud, a user can access different sub-systems integrated in the platform, including telescope time application, data search and access, cloud services (virtual machines, VOSpace, etc.), and public projects . The CSTCloud passport is a pair comprising an email address and a password that identifies a user in order to access services on the China Science \& Technology Network, and it is widely used by researchers at the Chinese Academy of Sciences.

The China-VO AstroCloud platform is considering accepting other third-party certifications from data centers and  science platforms such as CANFAR Arcade and NOAO Data Lab, as well as popular social media such as WeChat, Facebook, and Twitter. This would be convenient for non-researchers and users who do not have CSTCloud passports.

\subsection{ Challenges brings by mobile internet }
In the era of the mobile internet and Internet of Things (IoT), most online services support users from both the PC end and mobile end. Currently, the VO and e-science platform are designed mainly for PC users. The VO community should also develop apps for mobile end users. Due to the popularity of the 5G high-speed mobile network and IoT, new data sources will be available from the mobile internet. VO aims to provide full life cycle management for astronomical data, so it is important to solve ``the last-kilometer’’ problem by connecting a broader range of users, including smart telescopes, to its services.

VO or the science platform should facilitate the integration of mobile internet, cloud computing, big data, and the IoT to provide better services to improve research activities and public education.

\section{Future Trends and Discussion}
The amount of astronomy data will increase greatly in the near future. Thus, science platforms are being developed to allow researchers to efficiently analyze big data sets. These science platforms should support data mining and machine learning, and they need to be developed based on advanced cyber-infrastructure with the support and active involvement of the astronomy community.

\subsection{Really Big Data}
Astronomy is entering an era of big data because numerous ground-based and space-based photometric, spectroscopic, and time-domain surveys are in progress. These surveys are producing vast amounts of observational data, which cannot be transmitted or downloaded from data centers to the computers of astronomers as before. Processing these big data will require very powerful computing resources and the solution could be cloud computing \citep{2019BAAS...51g..55S}. Thus, all of the large survey data sets will be stored in the cloud and analyzed online, thereby changing the science research model from offline to online.

Several astronomical science platforms have been developed
     and running for some years, e.g., the NOAO Data Lab \citep{2014SPIE.9149E..1TF, 2019ASPC..523..233F}, CANFAR Arcade \citep{2019ASPC..523..277M}, and SciServer \citep{2017AAS...22923615R, 2019ASPC..521..749R}.

The China-VO AstroCloud is one of the earliest science platforms and it provides cyber-infrastructure for astronomy research. However, due to increases in the volume of data and technological development, China-VO plans to update the existing platform to provide better services for data archiving and management, data processing, and analysis under the umbrella of the newly established National Astronomical Data Center of China (NADC). In June 2019,  Ministry of Science and Technology of the People's Republic of China announced the first twenty national science data centers and NADC is one of them. 

\subsection{Data Mining and Machine Learning}
Astronomers have investigated the usage of cloud computing to help them process massive amounts of data, but simply offering virtual machines in the cloud is not sufficient for users. A key feature of a science platform is that it should provide an interactive and reproducible online data processing and analysis environment.

Most science platforms in astronomy and other disciplines employ a similar architecture and technologies to provide an interactive data analysis environment. JupyterHub\footnote{https://jupyter.org/hub} with JupyterLab\footnote{https://jupyterlab.readthedocs.io/en/latest/} are used as an interface for exploratory data mining and analysis and it has been investigated and adopted by several science platforms in the astronomy community, e.g in LSST \footnote{https://sqr-018.lsst.io/}. The interactive environment is generally deployed using container techniques (e.g., docker) , as in the LSST science platform.

Machine learning is used increasingly widely for astronomical data analysis. A science platform should provide a suitable environment for data exploration and machine learning. This is a common request for data analysis but an astronomical science platform should be tailored to its specific domain.  For example, the platform should enable easy filtering and preprocessing of astronomical data for machine learning algorithms.  General-purpose machine learning platforms like PAI developed by Alibaba Cloud cannot deal with special astronomical data formats, such as FITS, HDF5  and VOTable files. Also, the platform should integrate commonly used models and tools in astronomy that can be used as a starting point for model selection and optimisation.

\subsection{Cyber-infrastructure and Community Involvement}
High quality science platforms and data services need to be built on advanced cyber-infrastructure, including a very high-speed backbone network service with adequate storage and computing capacities. Take FAST as an example, about 1TB of data needs to be transferred from FAST site to NADC per hour, which requires advanced cyber-infrastructure with enough bandwidth to secure a good user experience. 

In addition, the support and  active involvement of the research community is essential for the development of domain-specific science platforms. The community could contribute to science requirements and developing consensus on data standards and policies regarding data access and sharing.  A successful science platform should have quite a number of active users. This requires the science platform to have advanced technology, professional user support and practical policies and regulations.This idea is reflected in the architecture of NADC.  

\begin{figure}[!h]
    \centering
    \includegraphics[width = \textwidth]{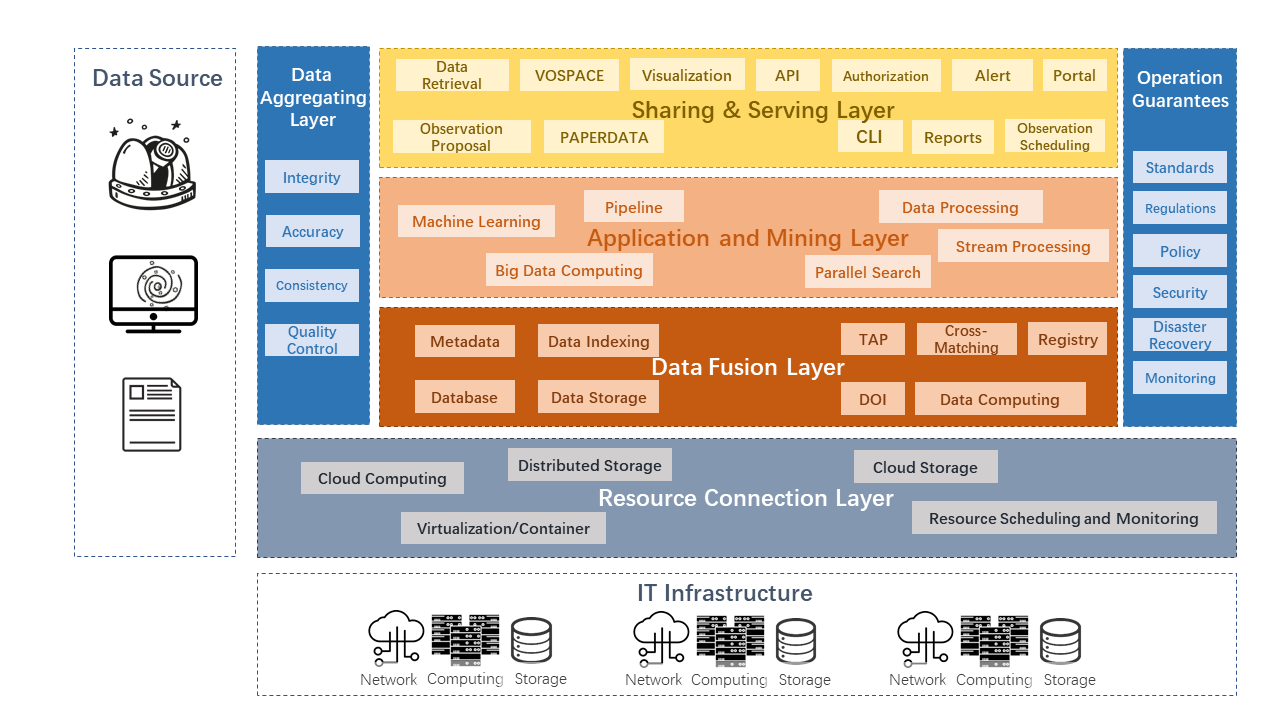}
    \caption{ Architecture of NADC.}
    \label{fig:NADC}
\end{figure}

In China, the NADC aims to fuse resources such as scientific data, literature, high-performance computing, software, and tools to provide one-stop services in terms of proposal submission and peer review, data archiving, and data sharing and usage in order to establish a physically distributed but logically unified science platform. As shown in Fig.\ref{fig:NADC}, the overall architecture of NADC mainly comprises resource connection layer,  data aggregating layer ,  data fusion layer, application and mining layer, and sharing and serving layer with operation guarantees.  In the data aggregating layer, astronomical data will be integrated into the platform as the Measures for the Management of Scientific Data required. The resource connection, data fusion, application and mining layer employ advanced computing technology and architecture to provide data services in the sharing and serving layer. The NADC will extend the current Astrocloud platform based on observational data resources, advanced cloud computing, big data, and VO technology to create a full life cycle management and open sharing platform for astronomical data. 

Alongside with a professional team and technical foundation, training is also important for the cultivation of  users. Science Platforms are bringing new research modes to researchers. Astronomer should embrace the changes. 
\section*{Acknowledgments}This work is supported by National Natural Science Foundation of China (NSFC)(11803055), the Joint Research Fund in Astronomy (U1731125, U1731243, U1931132) under cooperative agreement between the NSFC and Chinese Academy of Sciences (CAS), the 13th Five-year Informatization Plan of Chinese Academy of Sciences (No. XXH13503-03-107). Data resources are supported by China National Astronomical Data Center (NADC) and Chinese Virtual Observatory (China-VO). This work is supported by Astronomical Big Data Joint Research Center, co-founded by National Astronomical Observatories, Chinese Academy of Sciences and Alibaba Cloud. 

\bibliography{sciencePlatform}
\end{document}